\documentclass[12pt,a4paper]{article}
\usepackage{graphicx,amssymb,bm,latexsym,amsmath,braket,wrapfig,cite,color}
\usepackage{float}
\makeatletter
\@addtoreset{equation}{section}

\makeatother

\newcommand{\bea}{\begin{eqnarray}}
\newcommand{\eea}{\end{eqnarray}}

\newcommand{\pa}{\partial}
\newcommand{\til}{\widetilde}
\newcommand{\nn}{\nonumber\\}
\newcommand{\p}[1]{(\ref{#1})}

\newcommand{\tr}{\mathrm{Tr}}

\begin{document}
\begin{center}
{\fontsize{20}{0pt}\selectfont\bf
Gravitational renormalization group solutions and paticular effective actions} \\

\end{center}

\medskip
\renewcommand{\thefootnote}{\fnsymbol{footnote}}

\begin{center}
Daiki Yamaguchi
\footnote{e-mail: daichanqg@gmail.com} \\

\medskip

\emph{${}^1$Department of Physics, Kindai University, Higashi-Osaka, Osaka 577-8502, Japan
\\
\medskip
${}^2$Research Institute for Science and Technology, \\
Kindai University, Higashi-Osaka, Osaka 577-8502, Japan
\\
\medskip
${}^3$Related with Research Institution of Science and its Departments, \\
Kyosan University, Kamikaga-Kyoto, Kyoto, Japan}

\end{center}
\medskip

\begin{center}
{\bf Abstract}
\end{center}
We study the functional renormalization group equation and its solutions of the gravity having the background matters. From the system equivalence eliminating vacuum divergence, we are confirmed to give Newton coupling. We also give the path integral partition function technique to normalize setups of quantum corrected actions down to Einstein systems. Briefly, Einstein effective action and Stravinski effective action are come to the mass dependence with the appropriate cosmological constant.

\renewcommand{\thefootnote}{\arabic{footnote}}

\section{Introduction}
\label{sec:1}
When we consider the quantum field theory on curved spacetimes, we encounter the problem of the loop divergence given with quantum corrections. There is an important problem to make a finite effective action in quantum gravity. If we consider renormalizations of gravity, a functional renormalization group equation is a hot topic to give an asymptotic safe program of gravitational and field renormalizations. This equation gives gravitational and matter running couplings, which is verified from ultraviolet (UV) down to infrared (IR) as an asymptotic feature.

The historical talks of a functional renormalization group is based on Wilson renormalization method \cite{WK75,WH73}, which suggested that the partition function needs to explain the cutoff momentum of beta functions going lowered in IR. Some transformations of partition functions modified renormalization group equations and revealed the bound of Higgs mass, renormalizability of $\phi^4$ theorem, etc. After these days, the 1PI green function was said to be useful for thinking loop quantum corrections to give an effective action. From these reasons, a effective average action (EAA) was introduced by a regulator function $R_k(-q^2)$ to regularize loop divergences. Then, performing the Legendre transformation found a functional renormalization group equation, which was loop expanded with the second order has the one loop effect of the quantum correction. See for \cite{BB01,BTW02}. This equation absorbs loop divergences by set of running couplings depending on cutoff momentum scales.

By the way, thinking about solutions of spacetimes, a black hole singularity is trivial problem to explain the evaporation of this black hole. Any black holes in general relativity couldn't collapse into as the center of spacetimes. Actually, the black hole singularity is known to be resolved by the change of the spacetime topology in classical gravity, which is called as the regular black hole. There is a quantum effect to resolve the singularity problem of classical black holes in recent days. An interesting idea for making regular black holes with a functional renormalization improvements was found in \cite{BR00,BR06,RT11,KS142}, that scale identifications for black holes developed in \cite{PS18}. These discussions were indicating us to make the center regular. These talks are quantum improvements of black holes.

In this paper, we review the functional renormalization group equation calculations of the heat Kernel expansions. We consider the gravitational renormalization with the functional renormalization group with the system equibalence which is non vacumm divergence. In this situation, Newton coupling is given confirmed expression. Then, we introduce the new cosmological constant like Einstein convension. The two contracts of the cosmological constants give the quantum corrected action as special situations. To define the path integral normalization technique of the partition function, we could have Einstein systems as Einstein effective action and Stravinski action. The total effective action gives the inflation mass expression with the cosmological constant.

This paper is organized as follows. In section 2, we review a functional renormalization group equation and renormalization of the gravitational correction with background matters used by a local heat kernel master formula \cite{DEP}. Then, we introduce gravitational and cosmological running couplings mentioned by theoretical point of view. We prepare quantum corrected actions with sets of the cosmological constants. After that, we define a partition function of path integral methods and lead a quantum corrected Einstein action coming down to classical Einstein-Hilbert or Starobinski actions with renormalized couplings ansatz. We decide that the mass dependence with the cosmological constant. As a final topic of section 3, we summerize our results and roles of the quantum effects and quantum corrections.

\section{Graviton and matters}
\label{sec:2}

In this section, we briefly review or prove calculations for a gravitational renormalization used by a functional renormalization group equation. At first, we prepare a functional method as the heat kernel expansion to give beta functions on a simple theoretical space, which is defined by sets of arbitral manifold $S(\mu)$ and matter bundle $V(\Phi(\phi,A_\mu,\psi,\bar{\psi},...))$. After that, we calculate to show beta functions of Newton constant and cosmological constant. We also give the vacuum counter term to solve the vacuum divergence of the effective action with the cutoff energy disapeared. The original talk is done by \cite{DEP}. The eliminated cosmological constant is introduced as new cosmological constant and coupling remake the effective actions with the constant and simple contract of couplings. These remade actions including cutoff energy divergences these are the special system setups normalized to the gravitational effective actions are Einstein gravitational system.

\subsection{Local heat kernel expansion and a simple theoretical space}
We prepare a theoretical space for giving a convenient discussion of renormalization groups. Considering a gravitational manifold $S(\mu)$ and a matter domain $V(\Phi)$ described by a theoretical field $\Phi(\phi,A_\mu,\psi,\bar{\psi},...)$, we could define a theoretical space as $S(\mu)\otimes V(\Phi)$. These sets are meaning that gravitational manifold and matter fields are connected by a curvature. This curvature is defined by covariant derivatives as $\bf{\Omega}_{\alpha\beta}=[D_{\alpha},D_{\beta}]$ on theoretical space. If we consider a simple gravitational case never including $D(\nabla)=\nabla$, the theoretical space is only effective for the Einstein gravity. From these conversations, we prepare the effective action on such a simple space as follows,
\bea
\Gamma_{k}=\frac{1}{2}\int d^d x\sqrt{g}\Phi(x)(-\nabla^2+\bm{U}+\omega)\Phi(x).
\label{tea}
\eea

On the other hand, there are several techniques for giving the renormalization of physical couplings. Fundamental renormalization techniques are Feynman rules and Wilson renormalization group. However, a functional renormalization group equation is one of useful regulations for loop divergences detected by quantum field corrections. This is called as FRGE or Wetterich equation, which is given by,
\bea
\pa_t \Gamma_k=\frac{1}{2}\tr\Big(\frac{\delta^2 \Gamma_k}{\delta\Phi\delta\Phi}+R_k\Big)^{-1}\frac{dR_k}{dt}.
\label{frge}
\eea
Where, we denote a regulator $R_k(z=-\nabla^2+\bm{U})$. $R_k(z)$ is the cutoff function which goes to zero asymptotically with increases of a cutoff energy $k^2$. Here, we defined $z$ as a eigenvalue of the Laplace operator. Selections of regulator functions are some types, however, an optimized cutoff function,
\bea
R_k(z)=(k^2-z)\theta(k^2-z).
\label{opct}
\eea
$z$ is necessary notation for explaining the quantum gravity on $S(\mu)$, because of Fourie transformation being not satisfied on spacetimes. As this reason, we use a hessian identity as follows,
\bea
H=-\nabla^2+\omega+\bm{U}.
\label{ths}
\eea
\p{ths} gives a amplitude of a transition generated by $\bra{\sqrt{z'}}\Phi(x)\ket{\sqrt{z}}$. Therefore, we consider a propagator as \p{tea} inserted into \p{frge}, which is given by,
\bea
\frac{d\Gamma_k}{dt}=\frac{1}{2}\tr h_k(z) \ , \ \Big(h_k(z)=\frac{\pa_t R_k(z)}{z+\omega+R_k(z)}\Big).
\label{pfrg}
\eea
$h_k(z)$ is the propagator of the functional renormalization group. To solve the Wetterich equation that is calling the functional renormalization group equation, we use Laplace transformation below,
\bea
h_k(z)=\int_{0}^{\infty}ds \til h_k(s)e^{-sz}.
\label{ltr}
\eea
\p{pfrg} and \p{ltr} lead a heat kernel expansion of Wetterich equation,
\bea
\frac{d\Gamma_k}{dt}=\frac{1}{2}\int_0^\infty ds\til h_k(s)K_s(z), \ and \ (K_s(z)=\tr e^{-sz}).
\eea
Remaining our work is setting formulae of the trace heat kernel expansion. There are several definitions of heat kernel expansions in mathematics. To consider the renormalization technique, we usually use the local heat kernel expansion as below,
\bea
K_s(z)=\frac{1}{(4\pi s)^{d/2}}\int d^d x\sqrt{g}\Big(\tr \bm{b}_0(z)+s\tr\bm{b}_2(z)+...\Big).
\eea
We knew $s$ is the Laplace parameter. $\bm{b}_{0,2,4,...}$ are called as heat kernel coefficients. These are defined as follows,
\bea
\tr\bm{b}_0=\tr\bm{1} \ , \ \tr\bm{b}_2=\tr\Big(\frac{R}{6}\bm{1}-\bm{U}\Big).
\eea
the local heat kernel expansion is studied in detail \cite{BV87,BV90,A903,A902}. The mathematical sight for the heat kernel is given in \cite{CZ13}.
One loop divergence in the equation level is occurred in the order $\tr\bm b_2$. We also define $Q_n[f]$ functional,
\bea
Q_n[f]=\int_0^\infty ds s^{-n}f(s)=\frac{1}{\Gamma(n)}\int_0^{\infty}dw w^{n-1}f(w),
\label{qfc}
\eea
which simplifies the renormalization equation. Final equal in \p{qfc} is a Mellin transformation of regular $Q_n[f]$ function.
Wetterich equation \p{pfrg} and local heat kernel expansion \p{ltr} lead a master equation of the functional renormalization group equation as follows,
\bea
\frac{d\Gamma_k}{dt}=\frac{1}{2(4\pi)^{d/2}}\int d^d x\sqrt{g}\Big(Q_{d/2}[h_k(z)]\tr\bm{b}_0(z)+Q_{d/2-1}[h_k(z)]\tr\bm{b}_{2}(z)+...\Big).
\label{mfrge}
\eea
Finally, we expand the effective action with function $\bm\Phi_j$ as follows,
\bea
\Gamma_k(z)=\lim_{j\to \infty}\sum_{j}\beta(k)\Phi'_j=\int d^d x \sqrt{g}\beta(k)\Phi'(x).
\eea
If we assume a gravitational beta function $\beta(k)$, replacement $\Phi'\to R$ gives beta function. Therefore, the one loop divergence of quantum correction is calculated by \p{mfrge}. Some additional notations are going to be suggested in following discussions.

\subsection{Gravitational renormalization coupled with fields of matters}
In previous subsection, we prepared formulae for verifying the quantum loop correction. Here, we set Einstein gravity and back ground matter, which will be explained concretely as following computations. Here, we verify the functional renormalization group equation with graviton having the background matters and its coupling with matter numbers. We also give the vacuum counter term make the vacuum divergence of the effective action. Finally we confirm Newton coupling is the value of the functional renormalization processes as the asymptotic free.

We define effective actions as,
\bea
\Gamma_k=\Gamma_{EH}+\Gamma_{gf}+\Gamma_{FP}+\Gamma_{matters}.
\eea
Where, we set matters as follows,
\bea
\Gamma_{matters}=\Gamma_{S}+\Gamma_{D}+\Gamma_{U(1)}
+\Gamma_{U(1)gf}+\Gamma_{U(1)gh}.
\eea
An effective action of the wick rotated Einstein-Hilbert gravity is defined by,
\bea
\Gamma_{EH}=\frac{1}{2\kappa^2}\int d^d x \sqrt{g}(2\Lambda-R) \ , \ \kappa^2=8\pi G.
\eea
Then, we use a gauge fixing,
\bea
\Gamma_{gf}=\frac{1}{4\kappa^2}\int d^d x \sqrt{\bar g}\chi_{\mu}\chi^{\mu} \ , \ \Big(\chi_{\mu}=\bar{\nabla}^{\nu}h_{\mu\nu}-\frac{1}{2}\bar{\nabla}_{\mu}h\Big).
\eea
This is called as De-donder gauge condition with Feynman gauge.
And so on, the gauge fixing breaks gauge degrees of freedom, nevertheless, a Faddeev-Popov ghosts recover its symmetry as,
\bea
\Gamma_{FP}=\int d^d x\sqrt{\bar{g}}\bar{C}_{\mu}(\delta_{\nu}{}^{\mu}(-\nabla^2)-\bar{R}^{\mu}{}_{\nu})C^{\nu}.
\eea
Effective actions of matters are defined by,
\bea
\Gamma_{S}&=&\frac{1}{2}\int d^d x\sqrt{\bar{g}}\bar g^{\mu\nu}\sum_{i=1}^{N_{S}}\bar\nabla_{\mu}\phi^{i}\bar\nabla_{\nu}\phi^{i},\\
\Gamma_{D}&=&\int d^d x\sqrt{\bar{g}}\sum_{i=1}^{N_D}\bar{\psi}^{i}i\gamma^{\mu}D_{\mu}\psi^{i} \ , \ \Big(D_\mu=\pa_{\mu}+\frac{1}{8}[\Gamma^a,\Gamma^b]\omega_{\mu}{}^{ab}\Big).
\\
\Gamma_{U(1)}&=&\frac{1}{4}\int d^d x\sqrt{\bar{g}}\sum_{i=1}^{N_V}\bar{g}^{\mu\nu}\bar{g}^{\alpha\beta}
F^i_{\mu\alpha}F^i_{\nu\beta},
\\
\Gamma_{U(1)gf}&=&\frac{1}{2}\int d^d x\sqrt{\bar{g}}\sum_{i=1}^{N_V}(\bar{g}^{\mu\nu}\bar{\nabla}_{\mu}A^{i}_{\nu})^2, \ 
\\
\Gamma_{U(1)gh}&=&\int d^d x\sqrt{\bar{g}}\sum_{i=1}^{N_V}\bar{c}_{i}(-\bar\nabla^2)c_i.
\eea
Be careful, matters are defined as background fields. This means that matter fields numbers are effective to graviton corrections.
We derive hessians of effective actions. The gauge fixed Einstein action is given by,
\bea
\Gamma_{EH+gf}=\frac{1}{4\kappa^2}\int d^d x\sqrt{\bar g}h_{\alpha\beta}H^{\alpha\beta}{}_{\mu\nu}h^{\mu\nu},
\eea
Continually, the hessian is,
\bea
H^{\alpha\beta}{}_{\mu\nu}=K^{\alpha\beta}{}_{\mu\nu}\bar{R}+\frac{1}{2}(\bar{g}^{\alpha\beta}\bar{R}_{\mu\nu}+\bar{g}_{\mu\nu}\bar{R}^{\alpha\beta})-
\bar{g}^{(\alpha}{}_{(\mu}\bar{R}^{\beta)}{}_{\nu)}\nn
-\bar{R}^{(\alpha}{}_{(\mu}{}^{\beta)}{}_{\nu)}+(-\nabla^2-2\Lambda)K^{\alpha\beta}{}_{\mu\nu}
=(-\bar{\nabla}^2-2\Lambda)K^{\alpha\beta}{}_{\mu\nu}+U^{\alpha\beta}{}_{\mu\nu}.
\label{ghs}
\eea
Where, we used formulae in appendix.A. The symbol $K^{\alpha\beta}{}_{\mu\nu}$ is Dewitt metric, which is defined as,
\bea
K^{\alpha\beta\mu\nu}=\frac{1}{4}(\bar{g}^{\alpha\mu}\bar{g}^{\beta\nu}+\bar{g}^{\alpha\nu}\bar{g}^{\beta\mu}
-\bar{g}^{\alpha\beta}\bar{g}^{\mu\nu}).
\eea
Inverse Dewitt metric is,
\bea
K^{-1}_{\alpha\beta\mu\nu}=\bar{g}_{\alpha\mu}\bar{g}_{\beta\nu}
+\bar{g}_{\alpha\nu}\bar{g}_{\beta\mu}-\frac{2}{d-2}\bar{g}_{\alpha\beta}\bar{g}_{\mu\nu},
\eea
with the condition $K^{\alpha\beta}{}_{\rho\sigma}K^{\rho\sigma}{}_{\mu\nu}=\delta^{\alpha\beta}{}_{\mu\nu}$.
As \p{ghs} including the Dewitt metric ahead of the Laplace operator, we replace new gravitational hessian as $H_{EH+gf}{}^{\mu\nu}{}_{\alpha\beta}=K^{-1}{}^{\mu\nu}{}_{\rho\sigma}H^{\rho\sigma}{}_{\alpha\beta}$,
\bea
H_{EH+gf}{}^{\mu\nu}{}_{\alpha\beta}=(-\bar\nabla^2-2\Lambda)\delta^{\mu\nu}{}_{\alpha\beta}+2U^{\mu\nu}{}_{\alpha\beta}-\frac{d-4}{d-2}\bar{g}^{\mu\nu}\Big(\bar{R}_{\alpha\beta}-\frac{1}{2}\bar{g}_{\alpha\beta}\bar{R}\Big).
\eea
Considering Faddeev-Popov hessian, partial integral of the odd Grassmann number leads the result below,
\bea
\Gamma_{FP}=\int d^d x\sqrt{\bar{g}}\bar{C}_{\mu}H_{FP}{}^{\mu}{}_{\nu}C^{\nu},
\eea
and also, the hessian of Faddeev-Popov action is,
\bea
H_{FP}{}^{\mu\nu}=\bar{g}^{\mu\nu}(-\bar\nabla^2)-\bar{R}^{\mu\nu}.
\eea

Matter hessians are computed easier than the case of graviton. We derive hessians of scalar, Dirac, and gauge photon respectively in following discussions. A hessian of the scalar action is decided such as,
\bea
\Gamma_{S}=\frac{N_S}{2}\int d^d x\sqrt{\bar{g}}\phi(x)H_S\phi(x) \ , \ H_S=-\nabla^2.
\eea
A dirac hessian is given by,
\bea
\Gamma_{D}=N_D\int d^d x\sqrt{\bar{g}}\bar{\psi}(x)H_D^{1/2}\psi(x) \ , \ H_D=-D^2+\frac{\bar{R}}{4}.
\eea
Where, we used a formula of a vielbein assumptions,
\bea
\gamma^{\mu}D_{\mu}\gamma^{\nu}D_\nu=D^2-\frac{\bar{R}}{4}.
\eea
We could write down a gauge fixed $U(1)$ action and its hessian as,
\bea
\Gamma_{U(1)+U(1)gf}=\frac{N_V}{2}\int d^d x\sqrt{\bar g}A_\mu H^{\mu\nu}_{U(1)+U(1)gh}A_\nu \ , \ H^{\mu\nu}_{U(1)+U(1)gf}=(-\bar\nabla^2)\bar g^{\mu\nu}+\bar{R}^{\mu\nu}.
\eea
Finally, we find a ghost hessian as below,
\bea
\Gamma_{U(1)gh}=N_V\int d^d x\sqrt{\bar{g}}\bar{c}H_{U(1)gh}c \ , \ H_{U(1)gh}=-\bar\nabla^2.
\eea
Now on we are standing at the beginning point for explaining a graviton renormalization group, because those hessians have been computed already.
From hessians and \p{opct}, we could list propagators as follows,
\bea
h_k^{EH+gf}(z)&=&\frac{2k^2}{k^2-2\Lambda}\theta(k^2-z),\label{gpr}\\
h_k^{S,D,U(1)+U(1)gf,FP,U(1)gf}(z)&=&2\theta(k^2-z).\label{prs}
\eea
Therefore, Q-functionals defined in \p{qfc} are list as below,
\bea
Q_{d/2}[h_k^{EH+gh}]&=&\frac{1}{\Gamma(d/2+1)}\frac{2k^d}{1-2\til\Lambda}\\
Q_{d/2}[h_k^{S,D,U(1)+U(1)gf,FP,U(1)gh}]&=&\frac{2k^d}{\Gamma(d/2+1)}.
\eea
Except for the gravitational case, we took the eigenvalue $z=-\bar\nabla^2$. 

Our work is to calculate heat kernel coefficients. To give appropriate coefficients, we must be careful about dimensions of fields. The number of the dimensions are list as $\frac{d(d+1)}{2}$(graviton), $d$($U(1)$ and Faddeev-Popov), $2^{d/2}$(Dirac), and $1$(scalar and $U(1)$ ghosts). From these reasons, heat kernel coefficients are computed as,
\bea
\tr\bm{b}_{0}^{EH+gf}=\frac{d(d+1)}{2} \ , \ \tr\bm{b}_2^{EH+gf}=\frac{-5d^2+7d}{12}\bar{R},
\\
\tr\bm{b}_0^{FP}=d \ , \ \tr\bm{b}_2^{FP}=\frac{d+6}{6}\bar{R},
\\
\tr\bm{b}_0^S=1 \ , \ \tr\bm{b}_2^{S}=\frac{\bar{R}}{6},
\\
\tr\bm{b}_0^D=2^{d/2} \ , \ \tr\bm{b}_2^D=-2^{d/2}\frac{\bar{R}}{12},
\\
\tr\bm{b}_0^{U(1)+U(1)gf}=d \ , \ \tr\bm{b}_2^{U(1)+U(1)gf}=\frac{d-6}{6}\bar{R},
\\
\tr\bm{b}_0^{gh}=1 \ , \ \tr\bm{b}_2^{gh}=\frac{\bar{R}}{6}. 
\eea
Therefore, we could expand \p{frge} with \p{gpr} and \p{prs} as follows,
\bea
\frac{d\Gamma_k}{dt}=\frac{1}{2}\tr h_k^{EH+gf}(z)-\tr h_k^{FP}(z)+\frac{1}{2}\tr h_k^{S}(z)\nn
-\frac{1}{2}\tr h_k^D(z)+\frac{1}{2}\tr h_k^{U(1)+U(1)gf}(z)-\tr h_k^{U(1)gh}(z).
\eea 
Straightforwardly, master equations \p{mfrge} are given by,
\bea
\frac{1}{2}\tr h_k^{EH+gf}&=&\frac{1}{2(4\pi)^{d/2}}\int d^d x\sqrt{\bar{g}}\Big[\frac{1}{\Gamma(d/2+1)}\frac{d(d+1)}{2}\frac{2k^{d+2}}{k^2-2\Lambda}\nn
&~&+\frac{1}{\Gamma(d/2)}\frac{-5d^2+7d}{12}\frac{2k^d}{k^2-2\Lambda}\bar{R}\Big],
\\
-\tr h_k^{FP}&=&-\frac{1}{(4\pi)^{d/2}}\int d^d x\sqrt{\bar g}\Big[\frac{2k^d}{\Gamma(d/2+1)}d+\frac{2k^{d-2}}{\Gamma(d/2)}\frac{d+6}{6}\bar{R}\Big],
\\
\frac{1}{2}\tr h_k^S&=&\frac{N_S}{2(4\pi)^{d/2}}\int d^d x \sqrt{\bar g}\Big[\frac{2k^d}{\Gamma(d/2+1)}+\frac{2k^{d-2}}{\Gamma(d/2)}\frac{\bar{R}}{6}\Big],
\\
-\frac{1}{2}\tr h_k^D&=&-\frac{N_D}{2(4\pi)^{d/2}}\int d^d x\sqrt{\bar g}\Big[\frac{2k^d}{\Gamma(d/2+1)}2^{d/2}+\frac{2k^{d-2}}{\Gamma(d/2)}\Big(-2^{d/2}\frac{\bar{R}}{12}\Big)\Big],
\\
\frac{1}{2}\tr h_k^{U(1)+U(1)gf}&=&\frac{N_V}{2(4\pi)^{d/2}}\int d^d x\sqrt{\bar g}\Big[
\frac{2k^d}{\Gamma(d/2+1)}d+\frac{2k^{d-2}}{\Gamma(d/2)}\frac{d-6}{6}\bar{R}\Big],
\\
-\tr h_k^{U(1)gh}&=&-\frac{N_V}{(4\pi)^{d/2}}\int d^d x\sqrt{\bar{g}}\Big[\frac{2k^d}{\Gamma(d/2+1)}+\frac{2k^{d-2}}{\Gamma(d/2)}\frac{\bar{R}}{6}\Big].
\eea
As a result, Wetterich equation reaches,
\bea
\frac{d\Gamma_k}{dt}=\frac{1}{2(4\pi)^{d/2}}\int d^d x\sqrt{\bar{g}}\Big[\frac{2k^d}{\Gamma(d/2+1)}\Big(\frac{k^2}{k^2-2\Lambda}\frac{d(d+1)}{2}-2d+N_S+N_D 2^{d/2}+N_V(d-2)\Big)\nn
+\frac{2k^{d-2}}{\Gamma(d/2)}\bar{R}\Big(\frac{k^2}{k^2-2\Lambda}\frac{-5d^2+7d}{12}-\frac{d+6}{3}+\frac{N_S}{6}+\frac{N_D}{12}2^{d/2}+\frac{d-8}{6}N_V\Big)\Big].~~~~~
\label{resmeq}
\eea
As beta functions are given by set of a renormalized effective action. Then, we assume an effective action with renormalized couplings as follows,
\bea
\Gamma_k=\int d^d x\sqrt{\bar{g}}\Big(\frac{\til\Lambda}{8\pi \til G}-\frac{\bar{R}}{16\pi \til G}k^{d-2}\Big).
\label{refc}
\eea
A derivative of \p{refc} is written,
\bea
\frac{d\Gamma_k}{dt}=\int d^d x\sqrt{\bar{g}}\Big[\frac{k^d}{8\pi\til G}\Big(-\frac{\til\Lambda}{\til G}\frac{d\til G}{dt}+\frac{d\til\Lambda}{dt}+\til\Lambda d\Big)+\frac{k^{d-2}}{16\pi\til G}\bar{R}\Big(-(d-2)+\frac{1}{\til G}\frac{d\til G}{dt}\Big)\Big].
\label{cam}
\eea
Where, we defined dimensionless couplings $\til G$ and $\til\Lambda$.
Comparing \p{resmeq} and \p{cam}, we find beta functions. Newton constant (gravitational coupling) is given as below,
\bea
\frac{d\til G}{dt}
=(d-2)\til G+\frac{16\pi\til G^2}{(4\pi)^{d/2}\Gamma(d/2)}\Big(\frac{1}{1-2\til\Lambda}\frac{-5d^2+7d}{12}\nn
-\frac{d+6}{3}+\frac{N_S}{6}+\frac{N_D}{12}2^{d/2}+\frac{d-8}{6}N_V\Big).
\label{dgc}
\eea
If we consider $\til{\Lambda}<<1$ in $d=4$, we take \p{dgc} into account for,
\bea
\frac{d\til G}{dt}=2\til G-\frac{\til G^2}{6\pi}(46-N_S-2N_D+4N_V).
\label{gcp}
\eea
When we consider the condition $\frac{d\til G}{dt}=0$, a gravitational fixed point is given by,
\bea
g_*=\Big(\frac{46-N_S-2N_D+4N_V}{12\pi}\Big)^{-1}=\omega^{-1}.
\label{gfp}
\eea
Where, $\omega$ is the inverse fixed point. The beta function is rewritten by $\omega$ or $g_*$. To determine the gravitational coupling \p{gcp} in $d=4$, we fix the cosmological constant $\til \Lambda=0$. Remaining problem is the vacuum divergence of the effective action. The effective action $\gamma_k$ is the non vacuum divergent action these are eliminated by the system energy equivalences. Therefore the detected effective action is $\gamma_k$,
\bea
\frac{d\gamma_k}{dt}=\frac{d}{dt}(\Gamma_k-\Gamma_{vac}).
\eea
We could write down the effective action as,
\bea
\gamma_k=-\int d^dx\sqrt{\bar g}\frac{\bar R}{16\pi\til G}k^{d-2}.
\eea
The derivative is,
\bea
\frac{d\gamma_k}{dt}=\int d^d x\sqrt{\bar g}\Big[\frac{k^{d-2}}{16\pi\til G}\bar R\Big(-(d-2)+\frac{1}{\til G}\frac{d\til G}{dt}\Big)\Big].
\eea
Newton coupling is resemble to \p{dgc} as,
\bea
\frac{d\til G}{dt}
=(d-2)\til G+\frac{16\pi\til G^2}{(4\pi)^{d/2}\Gamma(d/2)}\Big(\frac{-5d^2+7d}{12}
-\frac{d+6}{3}+\frac{N_S}{6}+\frac{N_D}{12}2^{d/2}+\frac{d-8}{6}N_V\Big).
\eea
In d=4, we reach the same result of \p{gcp}. The gravitational fixed point is \p{gfp}. Our gravitational fixed point ensure the region $0<g_*<1$ with the minimal matter numbers. Newton coupling called as the gravitational coupling is given by,
\bea
\til G(k)=\frac{g_*G_0k^2}{g_*+ G_0k^2}, \ \ \ g_*=\Big(\frac{46-N_S-2N_D+4N_V}{12\pi}\Big)^{-1}.
\label{4gc}
\eea
The dimensionful Newton coupling is written as,
\bea
G(k)=\frac{g_* G_0}{g_* +G_0 k^2}.
\label{nc}
\eea
This Newton coupling behaives the asymptotic free scenario with the cutoff momentum $k$. We are care for the cutoff region $k[0,\infty)$. This means that Newton coupling goes from $G_0 (k=0)$ down towards $0^+ (k\to\infty)$.

\subsection{Special action setups and normalizations}
We derived Newton coupling in $d=4$ as \p{nc}. The effective action $\gamma_k$ is given with this Newton coupling. $\gamma_k$ includes the divergence of the equation scale $k^2$. Here we consider the special setups of $\gamma_k$ with the renewed cosmological couplings. Actions are normalized to Einstein system gravity from appropriate functional elimination methods, however, those are having cutoff vacuum also equation divergences. 

The vacuum divergence is eliminated of the system selection. Newton coupling satisfies the beta function generally,
\bea
\frac{d\til G}{dt}=2\Big(\til G-\frac{1}{g_*}\til G^2\Big).
\label{gbc}
\eea
The cosmological constant is easily introduced two methods. It is same as the cosideration with Einstein, the cosmological constant is given by the hand crafted properties as,
\bea
\Lambda_0>0, \ \ \textrm{or} \ \ \Lambda(k) G(k)=\Lambda_0 G_0 \ \to \ \Lambda(k)=G_0\frac{\Lambda_0}{G(k)}=\Lambda_0\Big(1+\frac{G_0}{g_*}k^2\Big).
\label{4cc}
\eea
The cosmological fixed point is directed by,
\bea
\til \Lambda(k)=\frac{\Lambda_0}{k^2}+\lambda_*=\frac{\Lambda_0}{k^2}+\Lambda_0\frac{G_0}{g_*}.
\eea
$\lambda_*$ is expressed as,
\bea
\lambda_*=\frac{\Lambda_0 G_0}{g_*}.
\eea
From this expression, $\lambda_*$ is in the region $0<\lambda_*$. Our processes are the two actions resetups with $(G(k),\Lambda_0)$ and $(G(k),\Lambda(k))$. the effective action $\gamma_k$ becomes the actions as,
\bea
S_{\Lambda_0}=\int d^4 x\sqrt{g}\frac{2\Lambda_0-R}{16\pi G(k)}, \ \ \ \ S_{\Lambda(k)}=\int d^4 x\sqrt{g}\frac{2\Lambda(k)-R}{16\pi G(k)}.
\label{qac}
\eea
We write down the special actions. The action $(G(k),\Lambda_0)$ is,
\bea
S_{\Lambda_0}=\frac{1}{16\pi G_0}\int d^4 x \sqrt{g}\Big[(2\Lambda_0-R)+\frac{G_0}{g_*}k^2(2\Lambda_0-R)\Big].
\label{eac}
\eea
The action $(G(k),\Lambda(k))$ is,
\bea
S_{\Lambda(k)}=\int d^4 x\sqrt{g}\frac{1}{16\pi G_0}\Big[(2\Lambda_0-R)+\frac{G_0}{g_*}(4\Lambda_0-R)k^2+2\Lambda_0\Big(\frac{G_0}{g_*}\Big)^2k^4\Big].
\label{eqac}
\eea
To understand $k$, we consider \p{eac} and \p{eqac} on flat spacetime. We find equations of motions from $\delta_k S(\eta;k)=0$,
\bea
k=0, \ \ \textrm{and} \ \ k\Big(1+\frac{G_0}{g_*}k^2\Big)=0 \ \to \ k=0.
\eea
$k\geq 0$ is the fundamental notation for the renormalization. As $k$ behaves as a field on the spacetime, this means that quantum field theory is thought with $k$. However, we don't know a canonical property of the scale momentum $k$. We could not quantize $k$ field canonically. In this case, path integral method is convenient for explaining the spacetime evolution. When we consider the quantum corrected Einstein action, we could write down a partition function of spacetime developments,
\bea
Z(g;\phi_k)=\int (d\phi_k)N[g]\exp[-S(g;\phi_k)].
\label{qpf}
\eea
Where we denote $\phi_k=k^n$ and $n=\bm{Z}^{+}$. We introduced normalize factor $N[g]$ to adjust the spacetime deformation. To prove a feature of the partition function, we compute an effective action with $(G(k),\Lambda_0)$ at first. The quantum corrected action is given by,
\bea
S(g;\phi^{G}_k)=\frac{1}{16\pi G_0}\int d^4 x \sqrt{g}\Big[(2\Lambda_0-R)+\frac{G_0}{g_*}(\phi_k^{G})^2(2\Lambda_0-R)\Big].
\label{eac}
\eea
Where, $R[g]$ doesn't depend on $k$. In this case the normalized factor is inverse property as the scalar cutoff scale integration. Such a property is written as,
\bea
N[g]=\Big[\int (dp)\exp\Big(-\frac{1}{16\pi G_0}\int d^4 x\sqrt{g}\frac{G_0}{g_*}p^2(2\Lambda_0-R)\Big)\Big]^{-1}.
\eea
Then, we find $\log Z(g;\phi_k^G)$,
\bea
\log Z(g;\phi_k^G)=-\frac{1}{16\pi G_0}\int d^4 x\sqrt{g}(2\Lambda_0-R).
\eea
The connected partition function is $W(g;\phi_k^G)$. Therefore, we find the effective action of $(G(k),\Lambda_0)$ as follows,
\bea
\Gamma^G (g)=\frac{1}{16\pi G_0}\int d^4 x\sqrt{g}(2\Lambda_0-R).
\eea
Continually, we consider couplings $(G(k),\Lambda(k))$ described by \p{eqac}. In this case, we set $\phi^{G,\Lambda}=k^2$. This ansatz leads \p{eqac} rewritten as,
\bea
S(g;\phi_k^{G,\Lambda})=\frac{1}{16\pi G_0}\int d^4 x\sqrt{g}\Big[2\Lambda_0\Big(\frac{G_0}{g_*}\Big)^2\Big(\phi_k^{G,\Lambda}+\frac{g_*}{4\Lambda_0 G_0}(4\Lambda_0-R)\Big)^2-\frac{R^2}{8\Lambda_0}\Big].
\eea
The normalize factor $N[g]$ is similarly computed as the inverse property of the integration $\phi_k^{G,\Lambda}$,
\bea
N[g]=\Big[\int(dp)\exp\Big(-\frac{1}{16\pi G_0}\int d^4 x\sqrt{g}2\Lambda_0\Big(\frac{G_0}{g_*}\Big)^2\Big(p+\frac{g_*}{4\Lambda_0 G_0}(4\Lambda_0-R)\Big)^2\Big)\Big]^{-1}.
\eea
Then, the connected partition function is given by,
\bea
\log Z(g;\phi_k^{G,\Lambda})=\frac{1}{128\pi G_0\Lambda_0}\int d^4 x\sqrt{g}R^2.
\eea
Therefore, we find the effective action as follows,
\bea
\Gamma^{G,\Lambda}(g)=\frac{-1}{128\pi G_0\Lambda_0}\int d^4 x\sqrt{g}R^2.
\eea
The actions are normalized into two important Einstein system effective actions. These are written in Lorentz frame,
\bea
(G(k),\Lambda_0) \ \to \ \Gamma_{EH}(g)&=&\frac{1}{16\pi G_0}\int d^4 x\sqrt{-g}(R-2\Lambda_0), \label{qeac}\\
(G(k),\Lambda(k)) \ \to \ \Gamma_{ST}&=&\frac{1}{128\pi G_0\Lambda_0}\int d^4 x\sqrt{-g}R^2.
\label{qsac}
\eea
Einstein effective action and Stravinski action are gained by normalization of the cutoff scale depending action setups. We also write down the full effective action,
\bea
\Gamma_{Full}=\frac{1}{16\pi G_0}\int d^4 x\sqrt{-g}\Big(R-2\Lambda_0+\frac{1}{8\Lambda_0}R^2\Big).
\label{eseq}
\eea
From \p{qsac}, Stravinski mass is respect to the value,
\bea
M^2=\frac{64}{3}\pi \Lambda_0.
\eea
We denote that the result of the full inflation action is meaning that the inflation mass property is depending on the cosmological constant. The mass behaives the logarism of the cosmological constant from $\Lambda_0=0$ towards $\Lambda_0=(Pos.finite)$. Inflation mass diverges with the cosmological constant goes to the positive infinity. 

\section{Results}
\label{sec:3}
The functional renormalization group equation is the modern technique for renormalize the gravitation with considering the one loop propagation having the cutoff energy function. Loop divergence is absorbed by running couplings as Newton constant and the cosmological constant. In this paper, we prepared Newton coupling having the cutoff energy scale of the equation divergence and appropriate system cutoff posession. Therefore, Newton coupling \p{nc} is briefly expressed also the fixed point \p{gfp} is given with matter numbers. We also consider the effective action renew setups with $(G(k),\Lambda_0)$ and $(G(k),\Lambda(k))$. The actions with situations $(G(k),\Lambda_0), ~(G(k),\Lambda(k))$ are normalized by the path integral mormalization methods towards Einstein effective action and Stravinski effective action. These effective actions are combined to the full effective action which is indication the inflation mass is the logarism of the positive cosmological constant.

We reviewed the graviton renormalization with the functional renormalization group (Wetterich) equation in section 2. This reveals gravitational and cosmological couplings derived from bata functions. These couplings are decided by exact calculations from the local heat kernel expansion of Wetterich equation. We also applied couplings as \p{4gc} and \p{4cc}, those properties are given by an appropriate counter term given from the system equiliblium of the vacuum divergence.

In last half section 2, we are reached the quantum corrected action these are the appropriate special setups of the cosmological constants. The normalization of the quantum corrected actions with situations $(G(k),\Lambda_0)$ and $(G(k),\Lambda(k))$ gives Einstein action and Stravinski action respectively. Here we notify that the normal factor is computed to eliminate the cutoff path integration part but comes the needless action term is resolved. This method open our ailes to give the effective actions such as Einstein Stravinski action. The full effective action \p{eseq} what is Einstein Stravinski action from the quantum corrected actions reveals the inflation mass is expressed by the cosmological constant.

To think about the results or analysis of the functional renormalization techniques, we finally have the quantum corrected actions. The quantum effects or corrections are having posibilities to birth the gravitational improvements of Einstein systems. Quantum corrected Einstein systems give the roots of the classical gravity as the induced parameter sights. The asymptotic scenario is also the quantum effects to the gravity could change the mass meanings.


\newpage

\appendix
\section{Background and fluctuations}
To consider the gravitational hessian, we should prepare perturbative expansions these are fluctuations. Quantum effects of the action is given by $h_{\mu\nu}^2$ terms. Thinking about the quantum effects of gravitation, we summarize here perturbative expansions from the background metric. We define a matric perturbative expansion as,
\bea
g_{\mu\nu}=\bar{g}_{\mu\nu}+h_{\mu\nu} \ , \ g^{\mu\nu}=\bar{g}^{\mu\nu}-h^{\mu\nu}+h^{\mu\alpha}h_{\alpha}{}^{\nu}.
\eea 
We agree with $g_{\mu\alpha}g^{\alpha\nu}=\delta_{\mu}{}^{\nu}$. Considering the spacetime volume $V=\int d^dx\sqrt{-g}$, we also give a metric determinant below,
\bea
\sqrt{-g}=\sqrt{-\bar{g}}\Big[1+\frac{h}{2}+\Big(\frac{1}{8}h^2-\frac{1}{4}h_{\mu\nu}h^{\mu\nu}\Big)\Big].
\eea

This metric expansion leads perturbative Christoffel symbols,
\bea
\Gamma^{\alpha}{}_{\mu\nu}=\bar{\Gamma}^{\alpha}{}_{\mu\nu}+\Gamma^{(1)\alpha}{}_{\mu\nu}+\Gamma^{(2)\alpha}{}_{\mu\nu}.
\eea
Then, we write down symbols of fluctuation parts,
\bea
\Gamma^{(1)\alpha}{}_{\mu\nu}=\frac{1}{2}(-\bar{\nabla}^{\alpha}{}h_{\mu\nu}+\bar{\nabla}_{\nu}h^{\alpha}{}_{\mu}+\bar{\nabla}_{\mu}h^{\alpha}{}_{\nu}),\\
\Gamma^{(2)\alpha}{}_{\mu\nu}=-\frac{1}{2}h^{\alpha\beta}(-\bar{\nabla}_{\beta}h_{\mu\nu}+\bar{\nabla}_{\nu}h_{\mu\beta}
+\bar{\nabla}_{\mu}h_{\nu\beta}).
\eea
Perturbative expansions of the Riemann tensor are,
\bea
R^{\mu}{}_{\nu\alpha\beta}=\bar{R}^{\mu}{}_{\nu\alpha\beta}+R^{(1)\mu}{}_{\nu\alpha\beta}+R^{(2)\mu}{}_{\nu\alpha\beta}.
\eea
We write down perturbative expansions as,
\bea
R^{(1)\mu}{}_{\nu\alpha\beta}=\frac{1}{2}(\bar{\nabla}_{\alpha}\bar{\nabla}_{\nu}h^{\mu}{}_{\beta}-\bar{\nabla}_{\alpha}\bar{\nabla}^{\mu}h_{\nu\beta}
-\bar{\nabla}_{\beta}\bar{\nabla}_{\nu}h^{\mu}{}_{\alpha}+\bar{\nabla}_{\beta}\bar{\nabla}^{\mu}h_{\nu\alpha})
+\frac{1}{2}\bar{R}_{\nu\rho\alpha\beta}h^{\mu\rho}+\frac{1}{2}\bar{R}^{\mu}{}_{\rho\alpha\beta}h^{\rho}{}_{\nu},~~~~~\\
R^{(2)\mu}{}_{\nu\alpha\beta}=-\frac{1}{2}h^{\mu\rho}\bar{\nabla}_{\alpha}(\bar{\nabla}_{\beta}h_{\nu\rho}+\bar{\nabla}_{\nu}h_{\beta\rho}
-\bar{\nabla}_{\rho}h_{\nu\beta})-\frac{1}{4}\bar{\nabla}_{\alpha}h^{\mu\rho}(\bar{\nabla}_{\beta}h_{\nu\rho}+\bar{\nabla}_{\nu}h_{\beta\rho}
-\bar{\nabla}_{\rho}h_{\nu\beta})\nn
+\frac{1}{4}\bar{\nabla}_{\rho}h^{\mu}{}_{\alpha}(\bar{\nabla}_{\beta}h^{\rho}{}_{\nu}+\bar{\nabla}_{\nu}h^{\rho}{}_{\beta}-\bar{\nabla}^{\rho}h_{\nu\beta})
-\frac{1}{4}\bar{\nabla}^{\mu}h_{\alpha\rho}(\bar{\nabla}_{\beta}h^{\rho}{}_{\nu}
+\bar{\nabla}_{\nu}h^{\rho}{}_{\beta}-\bar{\nabla}^{\rho}h_{\nu\beta})-(\alpha\leftrightarrow \beta).~~~~~
\eea
Then, fluctuations of Ricci tensor are contracted as $R_{\mu\nu}=R^{\rho}{}_{\mu\rho\nu}$ below,
\bea
R^{(1)}{}_{\mu\nu}=-\frac{1}{2}(\bar{\nabla}_{\mu}\bar{\nabla}_{\nu}h-\bar{\nabla}_{\mu}\bar{\nabla}_{\alpha}h^{\alpha}{}_{\nu}-\bar{\nabla}_{\nu}\bar{\nabla}_{\alpha}h^{\alpha}{}_{\mu}+\bar{\nabla}^2h_{\mu\nu})\nn
-\bar{R}_{\alpha\mu\beta\nu}h^{\alpha\beta}+\frac{1}{2}\bar{R}_{\mu\alpha}h^{\alpha}{}_{\nu}+\frac{1}{2}\bar{R}_{\nu\alpha}h^{\alpha}{}_{\mu},\\
R^{(2)}_{\mu\nu}=\frac{1}{2}\bar{\nabla}_{\mu}(h^{\alpha\beta}\bar{\nabla}_{\nu}h_{\alpha\beta})-\frac{1}{2}\bar{\nabla}_{\alpha}[h^{\alpha\beta}(\bar{\nabla}_{\mu}h_{\nu\beta}+\bar{\nabla}_{\nu}h_{\mu\beta}
-\bar{\nabla}_{\beta}h_{\mu\nu})]\nn
-\frac{1}{4}(\bar{\nabla}_{\mu}h^{\beta}{}_{\alpha}+\bar{\nabla}_{\alpha}h^{\beta}{}_{\mu}-\bar{\nabla}^{\beta}h_{\alpha\mu})(\bar{\nabla}_{\beta}h^{\alpha}{}_{\nu}+\bar{\nabla}_{\nu}h^{\alpha}{}_{\beta}-\bar{\nabla}^{\alpha}h_{\beta\nu})\nn
+\frac{1}{4}\bar{\nabla}_{\alpha}h(\bar{\nabla}_{\mu}h^{\alpha}{}_{\nu}+\bar{\nabla}_{\nu}h^{\alpha}{}_{\mu}-\bar{\nabla}^{\alpha}h_{\mu\nu}).
\eea
The contraction of the Ricci tensor leads scalars,
\bea
R^{(1)}=\bar{\nabla}_{\mu}\bar{\nabla}_{\nu}h^{\mu\nu}-\bar{\nabla}^2h
-\bar{R}_{\mu\nu}h^{\mu\nu},
\\
R^{(2)}=\frac{3}{4}\bar{\nabla}_{\alpha}h_{\mu\nu}\bar{\nabla}^{\alpha}h^{\mu\nu}+h_{\mu\nu}
\bar{\nabla}^2 h^{\mu\nu}-\bar{\nabla}_\alpha h^{\alpha}{}_{\mu}\bar{\nabla}_{\beta}h^{\beta\mu}+\bar{\nabla}_{\alpha}h^{\alpha}{}_{\mu}\bar{\nabla}^{\mu}h-2h_{\mu\nu}\bar{\nabla}^{\mu}\bar{\nabla}_{\alpha}h^{\alpha\nu}
\nn
+h_{\mu\nu}\bar{\nabla}^{\mu}\bar{\nabla}^{\nu}h-\frac{1}{2}\bar{\nabla}_{\mu}h_{\nu\alpha}\bar{\nabla}^{\alpha}h^{\mu\nu}-\frac{1}{4}\bar{\nabla}_{\mu}h\bar{\nabla}^{\mu}h
+\bar{R}_{\mu\nu\alpha\beta}h^{\mu\alpha}h^{\nu\beta}.
\eea
Here, we neglect the total derivative terms. Then we rewrite $R^{(2)}$ as below,
\bea
R^{(2)}=\frac{1}{4}(h_{\mu\nu}\bar{\nabla}^2h^{\mu\nu}+h\bar{\nabla}^2h
+2(\bar{\nabla}^{\alpha}h_{\mu\alpha})^2
+2\bar{R}_{\mu\nu}h^{\mu\alpha}h^{\nu}{}_{\alpha}+2\bar{R}_{\mu\nu\alpha\beta}h^{\mu\alpha}h^{\nu\beta}).
\eea
These relations are used to compute the gravitational hessian. Hessian is given by the fluctuation $R^{(2)}$, this means the inverse propagator of the graviton. Once we know the loop dependence of the gravitational propagator, we could verify processes of the gravitational functional renormalization with appropriate cutoff functions.

\section{Covariant derivative of Dirac field}
One of important problems are a treatment of covariant derivative of Dirac fields. In these contents, we used the vielbein assumption and the specific definition of the covariant derivative of Dirac fields. After discussions, we show definitions of covariant derivatives in this section. Spinor is not invariant under any transformations. In this case, to consider the local flat feature of transformations on Dirac field, we prepare vielbein as follows,
\bea
b^i{}_{\mu}b_{i\nu}=g_{\mu\nu} \ , \ b_{i\mu}b^{\mu}{}_{j}=\eta_{ij}.
\eea
The most simple way to see the local feature, we also consider the infinitesimal Loretz transformation,
\bea
x'^{i}=\Lambda^{i}{}_{j}x^j=\Big(1-\frac{1}{2}\epsilon^{ab}S_{ab}\Big)^{i}{}_{j}x^{j}.
\eea
$S_{ab}$ parameter is defined as,
\bea
(S_{ab})^{i}{}_{j}=\delta^{i}{}_{a}\eta_{bj}-\delta^{i}{}_{b}\eta_{aj}.
\eea
The commutator relation of $S_{ab}$ is important to give a relation as follows,
\bea
[S_{ij},S_{ab}]=(\eta_{ia}S_{jb}-\eta_{ja}S_{ib}-\eta_{ib}S_{ja}+\eta_{jb}S_{ia}).
\eea
When we consider the Lorentz transformation, we define a covariant derivative as follows,
\bea
D_\mu \bm{\Phi}=\pa_\mu \bm{\Phi}+\frac{1}{2}\pa_{\mu}\omega^{ij}S_{ij}\bm{\Phi}.
\eea
Where the field $\bm{\Phi}$ is on theoretical space. Remember $S_{ij}=-S_{ji}$, parameter $\omega^{ij}$ satisfies $\pa_{\mu}\omega^{ij}=-\pa_\mu \omega^{ji}$. The infinitesimal Lorentz transformation gives changes of parameters as,
\bea
\pa_{\mu}\omega^{ij}\to \pa_{\mu} \omega^{ij}+\pa_{\mu} \delta\omega^{ij} \ , \ 
\pa_{\mu} \delta\omega^{ij}=\epsilon^{i}{}_{a}\pa_{\mu}\omega^{aj}+\pa_{\mu}\epsilon^{ij}.
\eea
Therefore, the derivation of the covariant derivative satisfies a relation,
\bea
\delta(D_{\mu}\bm{\Phi})=\frac{1}{2}\epsilon^{ij}S_{ij}D_\mu \bm{\Phi}.
\eea
This $D_\mu$ is different from the gravitational covariant derivative $\nabla_{\mu}$. $D_\mu$ is a covariant derivative on the local Lorentz frame. Using $d_\mu$, a vielbein satisfies as follows,
\bea
D_\mu b_{k\nu}=\pa_{\mu}b_{k\nu}+\pa_{\mu}\omega_{kj}b^{j}{}_{\nu}.
\eea
$b_{a\mu}$ is invariant under the general coordinate transformation. Then, the covariant derivative is defined by,
\bea
\mathcal{D}_{\mu}b_{k\nu}=D_\mu b_{k\nu}-\Gamma^{\alpha}{}_{\mu\nu}b_{k\alpha},
\eea
which including the gravitational term as the affine connection (Christoffel symbol). We also assume vielbein assumption as,
\bea
\mathcal{D}_{\alpha}g_{\mu\nu}=0 \ \leftrightarrow \ \mathcal{D}_{\alpha}b_{k\mu}=0.
\eea
Then, we find the relation of the covariant derivative,
\bea
D_\mu b_{k\nu}=\Gamma^{\alpha}{}_{\mu\nu}b_{k\alpha}.
\eea
The vielbein assumption is the idea for revealing a curvature on local Lorentz frame. To do so, we calculate the commutator as follows,
\bea
[D_\mu,D_\nu]\bm{\Phi}=\Big(\pa_{[\mu}\pa_{\nu]}\omega^{ij}S_{ij}+\frac{1}{2}\pa_{[\mu}\omega^{kl}\pa_{\nu]}\omega^{ij}S_{kl}S_{ij}\Big)\bm{\Phi}.
\eea
The second term is symmetric of indices $ij\leftrightarrow kl$. When we define a curvature $R^{ij}{}_{\mu\nu}$, 
\bea
[D_\mu,D_{\nu}]\bm{\Phi}=\frac{1}{2}R^{ij}{}_{\mu\nu}S_{ij}\bm{\Phi},
\eea
we find the curvature expression as follows,
\bea
R^{ij}{}_{\mu\nu}=2(\pa_{[\mu}\pa_{\nu]}\omega^{ij}+\pa_{[\mu}\omega^{i}{}_{k}\pa_{\nu]}\omega^{kj}).
\eea
Where, we remind $b^{a\mu}b_{a}{}^{\nu}=g^{\mu\nu}$. Therefore, we find a curvature as follows,
\bea
R^{\mu}{}_{\alpha\beta\gamma}=b^{\mu}{}_{i}b_{\alpha j}R^{ij}{}_{\beta\gamma}.
\eea

Finally, we consider the covariant derivative corresponding to Dirac derivative. In this case, we could calculate the Laplace operator as below,
\bea
(\Gamma^{\mu}D_{\mu})^2\psi&=&\Gamma^{\mu}D_\mu \Gamma^{\nu}D_\nu \psi \nn
&=&(\eta^{\mu\nu}+\gamma^{\mu\nu})D_\mu D_\nu\psi \nn
&=&\Big(D^2+\frac{\gamma^{\mu\nu}}{4}R^{ij}{}_{\mu\nu}S_{ij}\Big)\psi \nn
&=&\Big(D^2+\frac{1}{8}\gamma^{\mu\nu}R_{\mu\nu\alpha\beta}\gamma^{\alpha\beta}\Big)\psi.
\eea
Where, $\gamma^{\mu\nu}$ is expanded by $\Gamma^\mu$. Therefore, we find the relation as follows,
\bea
\frac{1}{8}\gamma^{\mu\nu}R_{\mu\nu\alpha\beta}\gamma^{\alpha\beta}&=&\frac{1}{8}\Gamma^{\mu}\Gamma^{\nu}\Gamma^{\alpha}\Gamma^{\beta}R_{\mu\nu\alpha\beta}\nn
&=&\frac{1}{8}\Big(g^{\mu\nu}\Gamma^{\alpha}-g^{\mu\alpha}\Gamma^{\nu}+g^{\nu\alpha}\Gamma^{\mu}+\gamma^{\mu\nu\alpha}\Big)
\Gamma^{\beta}R_{\mu\nu\alpha\beta}=-\frac{R}{4}.
\eea
As a result, we find the Laplace operator as follows,
\bea
(\Gamma^{\mu}D_{\mu})^2\psi=\Big(D^2-\frac{R}{4}\Big)\psi.
\eea
We use this formula to compute a quantum theory on background Dirac field. If we consider the gravitational Laplace covariant derivative, we know $D_\mu$ including the gravitational metric derivative with the slight changes.

\end{document}